\documentclass{ametsocV6.1}

\newcommand{\defn}{\ensuremath{\stackrel{\textrm{def}}{=}}}
\renewcommand{\equiv}{\defn}

\newcommand{\lnflag}{\nolinenumbers} 
\definecolor{green}{rgb}{0.0, 0.47, 0.44}
\definecolor{purple}{rgb}{0.5, 0.0, 0.5}
\newcommand{\cpurple}[1]{{#1}}
\newcommand{\cblue}[1]{{#1}}
\newcommand{\cgreen}[1]{{#1}}
\usepackage{url} 
\usepackage{lineno}
\usepackage{amsmath}
\usepackage{comment}
\usepackage{soul}
\lnflag

\title{Surface to Seafloor: A Generative AI Framework for Decoding the Ocean Interior State}
\authors{Andre N. Souza\aff{a}\correspondingauthor{\lnflag Andre Nogueira de Souza, andrenogueirasouza@gmail.com}, Simone Silvestri\aff{a}, Katherine Deck\aff{b}, Tobias Bischoff\aff{c}, Glenn R. Flierl\aff{a}, and Raffaele Ferrari\aff{a}}

\affiliation{
\lnflag
\aff{a}{Massachusetts Institute of Technology, Cambridge, MA, USA},
\newline
\aff{b}{California Institute of Technology, Pasadena, CA, USA},
\newline
\aff{c}{Aeolus Labs, San Francisco, CA, USA}
}

\abstract{\lnflag  Understanding subsurface ocean dynamics is essential for quantifying oceanic heat and mass transport, but direct observations at depth remain sparse due to logistical and technological constraints. In contrast, satellite missions provide rich surface datasets—such as sea surface height, temperature, and salinity—that offer indirect but potentially powerful constraints on the ocean interior. Here, we present a probabilistic framework based on score-based diffusion models to reconstruct three-dimensional subsurface velocity and buoyancy fields, including the energetic ocean eddy field, from surface observations. Using a 15-level primitive equation simulation of an idealized double-gyre system, we evaluate the skill of the model in inferring the mean circulation and the mesoscale variability at depth under varying levels of surface information. We find that the generative model successfully recovers key dynamical structures and provides physically meaningful uncertainty estimates, with predictive skill diminishing systematically as the surface resolution decreases or the inference depth increases. These results demonstrate the potential of generative approaches for ocean state estimation and uncertainty quantification, particularly in regimes where traditional deterministic methods are underconstrained or ill-posed.}
\begin{document}
\lnflag

\maketitle

%
%
%
%
%
%

%


\section{Introduction}
Inferring subsurface ocean dynamics from surface observations remains a fundamental challenge in physical oceanography. The ocean interior plays a crucial role in regulating Earth's climate by transporting heat, carbon, and other tracers, yet direct measurements at depth are sparse due to logistical and technological limitations. In contrast, satellite missions have provided decades of continuous, high-resolution observations of sea surface height (SSH) and temperature and large-scale salinity observations, offering a vastly richer dataset than what is available below the surface \citep{see_sea}. Theoretical and numerical studies suggest that surface anomalies—such as sea surface height—strongly correlate with subsurface structures at mesoscale and submesoscale scales, meaning surface data could provide meaningful constraints on the ocean interior \citep{Hurlburt1986, CooperHaines1996}. However, existing methods for reconstructing subsurface states from surface observations are often limited by strong dynamical assumptions or high computational costs. 

This study focuses on reconstructing interior velocity and temperature fields, with a particular emphasis on their correlations, which are essential for estimating mass and heat transport throughout the ocean. Mesoscale eddies, which account for 20–50\% of total heat transport \citep{su2018, saenko2018}, are particularly difficult to infer from surface observations alone. A deterministic reconstruction, which attempts to match a single ``true" state, may fail to capture these dynamically significant features. Instead, we take a probabilistic approach, where a generative model reconstructs a distribution of plausible subsurface states statistically consistent with available surface data. This methodology ensures that eddy statistics, such as their variance, spatial correlations, and contributions to heat transport, are preserved without erasing dynamically important details.

Several physically motivated methods have been developed to infer subsurface ocean states from surface data, each with distinct advantages and limitations. \cgreen{The Surface Quasi-Geostrophic (SQG) framework has been used for subsurface reconstruction. The theory works well when the inherent assumptions in QG theory are satisfied and there exists information about the vertical stratification \citep{Held_Pierrehumbert_Garner_Swanson_1995, DynamicsoftheUpperOceanicLayersinTermsofSurfaceQuasigeostrophyTheory, ReconstructingtheOceansInteriorfromSurfaceData, ReconstructingtheOceanInteriorfromHighResolutionSeaSurfaceInformation}. Theoretical extensions, such as the SQG model with a multivariate Empirical Orthogonal Function have been shown to improve upon standard SQG by accurately reconstructing both surface- and subsurface-intensified eddies, particularly in regions with complex vertical stratification like the Kuroshio Extension and the Peru-Chile upwelling system~\citep{SQG-mEOF-R}. However, recent satellite missions such as the Surface Water and Ocean Topography (SWOT) Mission \citep{swot} achieve higher resolution than traditional altimeters (down to 10 km) and resolve scales where the the QG approximations are no longer valid, necessitating additional theoretical development or alternative approaches.  } 

Theoretical limitations can be overcome in the presence of data-abundance. In particular, machine learning offers a data-driven alternative to augment or completely replace physically constrained methods for ocean state estimation. Its applications range from model emulation \citep{samudra2024} to data assimilation \citep{CHAMPENOIS2024134026}, with increasing success in leveraging satellite observations. Linear models have demonstrated the statistical predictability of large-scale ocean currents, such as the Atlantic Meridional Overturning Circulation (AMOC), using satellite-derived data \citep{cromwell2007towards, delsole2022reconstructing}. More recently, feed-forward neural networks (FNNs) and convolutional neural networks (CNNs) have been applied to reconstruct meridional overturning circulations from ocean bottom pressure, zonal wind stress, and sea surface properties \citep{solodoch2023machine, meng2024circumpolar}. Additionally, studies incorporating satellite and Argo float data have demonstrated skill in inferring velocity at depth \citep{chapman2017reconstruction}, while CNNs trained on quasi-geostrophic (QG) simulations have successfully mapped surface stream function anomalies to subsurface flow fields in idealized settings \citep{Bolton2018}. Complementary efforts using QG theory \citep{Manucharyan2021} and data assimilation techniques \citep{Manucharyan2023} have refined our understanding of large-scale ocean dynamics. Meanwhile, operationally constrained models integrate observational data with numerical simulations, leading to more accurate reconstructions of subsurface ocean conditions \citep{Manucharyan2023_oc}.

While these studies have demonstrated that satellite-derived surface information can be used to infer the ocean interior, most have relied on supervised learning, where a model is trained to minimize residuals against a single ``true" state. However, a given surface ocean state does not deterministically correspond to a unique deep ocean state. A more physically sound approach requires probabilistic modeling, which allows for generating an ensemble of plausible subsurface states constrained by surface observations, while preserving the uncertainties and variability inherent in ocean dynamics. \cpurple{Probabilistic methods can be considered an extension of supervised learning methods as they can often reproduce the inference from such methods, as a subset of their capabilities. For example, minimizing the root-mean-square error of a probabilistic output results in the mean of the underlying distribution which is the target of supervised learning methods (barring discrepancies in training and architecture choices).}

Importantly, probabilistic modeling allows for uncertainty quantification (UQ) \citep{ECCO, kpp_uq}: not only can samples from the inferred distribution be explored, but these samples also provide estimates of plausible subsurface fields when surface information is insufficient to constrain the output fully. Generative modeling offers a flexible approach to sampling from such distributions, with score-based diffusion models emerging as a particularly effective tool for high-dimensional statistical inference. Generative-adversarial models (GANs) have also been applied to predict ocean temperature profiles from sea surface temperature \citep{meng2021physics, zhang2023deriving}, but these works did not leverage the potential for uncertainty quantification. The study in \cite{martin2025genda} leveraged score-based generative methods in the context of ocean surface data assimilation but again, they did not focus on uncertainty quantification; however, there have been recent efforts at examining uncertainty quantification in the presence of sparse ocean-like measurements in a quasi-geostrophic model \cite{babu2025guidedunconditionalconditionalgenerative}.  Score-based generative methods in particular have also been explored when addressing extreme event sampling as well \cite{STAMATELOPOULOS2025117589}. Beyond conditional sampling, score-based generative models have demonstrated utility in diverse applications, including calculating response functions \citep{Giorgini2024Response, giorgini2025predictingforcedresponsesprobability}, estimating the dimensionality of data manifolds \citep{stanczuk2023diffusion}, stabilizing the dynamical evolution neural network models \citep{pedersen2025thermalizerstableautoregressiveneural}, parameter calibration \citep{giorgini2025statisticalparametercalibrationgeneralized}, data-assimilation \citep{manshausen2025generativedataassimilationsparse}, or devising stochastic analogs to chaotic equations \citep{giorgini2025kgmmkmeansclusteringapproach, giorgini2025datadrivendecompositionconservativenonconservative, giorgini2025reducedordermodelingcyclostationarytime}. 

This study develops a probabilistic generative approach to infer the ocean interior state conditional on observable surface variables, such as sea surface height. By framing subsurface inference as a conditional sampling problem, we aim to quantify the variability and uncertainty of ocean interior states while leveraging the wealth of information encoded in surface observations. Rather than simply predicting a single deterministic state of ocean currents from the surface, our objective is to determine to what extent observations of the ocean surface constrain the ocean interior state while ensuring that the whole subsurface ocean state, including correlations amongst different variables, are properly captured. As we will show, a well-trained model can even predict the extent to which it is wrong in its predictions. 

With these aims in mind, the rest of the paper is organized as follows. In Section \ref{s:mcd}, we provide the physical context for the study, introduce the numerical simulation that we use as a testbed for the generative method, and describe the generative modeling framework. Results are presented in Section \ref{s:results}, where we apply the generative model to the ocean simulation and assess its predictive skill in reconstructing subsurface velocities and temperature alongside their variability as a function of sea surface height. We also repeat the exercise with spatially filtered sea surface heights to understand how less informative surface conditions alter predictions. Section \ref{s:end} concludes and discusses further extensions of this work. 

\section{Numerical Simulations and Generative AI Model}
\label{s:mcd}

In the following subsections, we describe the details of the ocean model output, the generative AI model, and training. In essence, we will use an ocean simulation to generate a time series of the ocean state, consisting of the sea surface height (SSH), velocities, and temperature at various vertical levels. This dataset then forms the backbone for training a neural network to determine plausible velocities and temperatures given a sea surface height. 

\subsection{Ocean Simulation}
\label{s:ocean}

 \cgreen{We illustrate the generative AI method with an idealized simulation rather than reanalysis data or a more realistic ocean setup for two primary reasons: 1) with an idealized study it is possible to generate (re-generate) as much training data as needed to test the methodology and 2) a more realistic ocean model introduces additional challenges like measurement uncertainty, data gaps, and nonstationary statistics. We believe that one must first demonstrate that the proposed methodology works in an idealized setting here before attempting it in a more complex scenario.   }

As with any idealized simulation, we focus on particularly salient features that generalize to more complex scenarios. Thus, the role of the simulation is not to capture realistic features of the ocean but rather to serve as a testing ground for whether or not statistical inference of the free surface to the ocean interior can be performed. With this in mind, we choose a baroclinic double gyre, similar to \citep{ANumericalModeloftheVentilatedThermocline}, as a representative ocean simulation for the methodology. The simulation generates circulation patterns consistent with those observed in the subtropical and subpolar gyres.
\textit{We will not run the ocean model to full statistical equilibrium} so that we can study both statistically stationary and non-stationary features, \citep{hasselman}. \cblue{We use a strict interpretation of ``full statistically-nonstationary'' to imply that ``all'' metrics of the simulation are converged. In our simulation only some statistics achieve stationarity during spinup, while other do not. The generative AI approach implicitly assumes stationarity of all statistics if it is to extrapolate future states from past ones. We will therefore note when the inference problem is applied to stationary versus non-stationary statistics.}

Using Oceananigans.jl \citep{Ramadhan2020, Silvestri2024_weno, wagner2023formulation,  Silvestri2024_meso, wagner2025oceananigans}, we solve the Boussinesq equations under the hydrostatic approximation. The dynamical core uses a nonlinear free surface and $z^\star$ coordinates. The prognostic variables are the horizontal velocities, $U$ and $V$, the SSH $\eta$, and the temperature $T$. A linear equation of state relates the buoyancy of seawater $B$ to the temperature, \citep{roquet}. The figures will refer to the Oceananigans simulation output as ``OcS'' for brevity. 


The idealized ocean domain extends from 15 to 75 degrees north in latitude and from 0 to 60 degrees in longitude with walls on all sides. The domain is 1800~m deep with a flat bottom. A zonal wind stress is applied at the surface, independent of longitude, with a cosine profile peaking eastward (positive) in the middle of the domain at a value of $0.1 \text{ N} \text{m}^{-2}$ and at a minimum and westward (i.e. negative) at the northern and southern walls. The surface temperature is restored to a linear profile in latitude ranging from $30$  C$^{\circ}$ on the southern boundary at $15^\circ$N, to $0$ C$^{\circ}$ at the northern boundary at $75^\circ$N. A quadratic bottom drag is imposed at the seafloor and convective adjustment is used for vertical mixing. The (dimensionless) quadratic drag coefficient is $10^{-3}$ and we use a horizontal diffusivity and viscosity of $2.5 \times 10^{2} \text{ m}^2 \text{s}^{-1}$, a background vertical diffusivity of $10^{-5} \text{ m}^2 \text{s}^{-1}$, and background vertical viscosity of $10^{-2} \text{ m}^2 \text{s}^{-1}$.

The domain is discretized into $256 \times 256 \times 15$ grid points, leading to a 25-kilometer resolution in latitude. The longitudonal resolution depends on the given latitude, and is calculated by multiplying the latitudonal resolution with the cosine of the latitude, which ranges from factors of $0.96$ to $0.26$. The vertical grid spacing varies from 50 meters at the surface to 180 meters at depth. The simulation is initialized with zero velocity, a flat sea surface, and uniform buoyancy and runs for $\sim 425$ years. The ocean circulation spins up in the first $\sim 100$ years, which are discarded in the following analysis. Once spun up, the solution is characterized by a double gyre circulation with convection along the northern boundary and a rich mesoscale eddy field.  However, the ocean depths have not yet reached statistical equilibrium and continues to evolve. 

\begin{figure*}[htbp]
\vskip 0.2in
\begin{center}
\centerline{\includegraphics[width=1.0\textwidth]{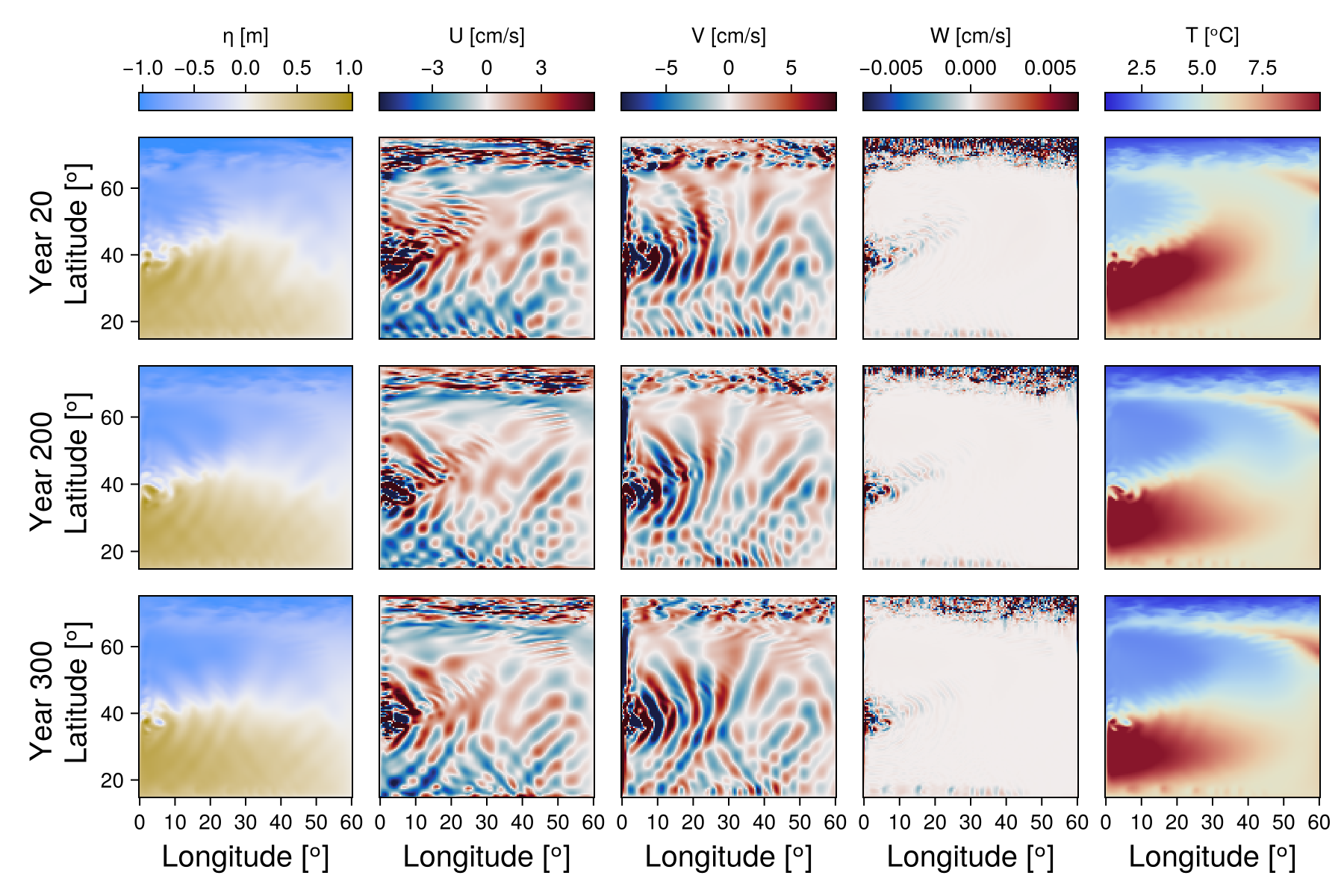}}
\caption{{Snapshots of the ocean state from a double-gyre simulation}. Each column is a different ocean state, and each row is a fixed moment in time. The first row consists of year 20, the middle row year 200, and the last row year 300. The columns are the sea surface height, the zonal velocity $U$, the meridional velocity $V$, the vertical velocity $w$, and the temperature $T$. The velocities and temperature are evaluated at a 400-meter depth. }
\label{fig:samples_in_time}
\end{center}
\vskip -0.2in
\end{figure*}

We show the resulting fields and their evolution in time in Figure \ref{fig:samples_in_time}. From left to right, each column shows the SSH $\eta$, the zonal $U$, meridional $V$, vertical velocity $W$, and temperature $T$ at a $400$ meter depth. Each row is a snapshot at years 20, 200, and 300. There is a cyclonic gyre in the northern half of the domain and an anti-cyclonic gyre in the lower half. The convection in the north half of the domain cools fluid parcels that sink and fill the abyss with cold water. The resulting simulation exhibits a non-trivial depth-dependent structure, deviating from equivalent barotropic behavior and includes a substantial ageostrophic component at all vertical levels.

The generative AI task of this manuscript is to generate plausible predictions for the ocean state $(U, V, W, T)$ at all depths, given the first column, the SSH $\eta$. We use the instantaneous SSH to infer plausible subsurface fields. We use this simulation as a proxy for performing ocean inference given satellite information and a model for the ocean interior. We only focus on the SSH since the imposed forcing constrains the present simulation's surface temperature field. In a more realistic setup, we would use more information. 

\begin{figure*}[htbp]
\vskip 0.2in
\begin{center}
\centerline{\includegraphics[width=\textwidth]{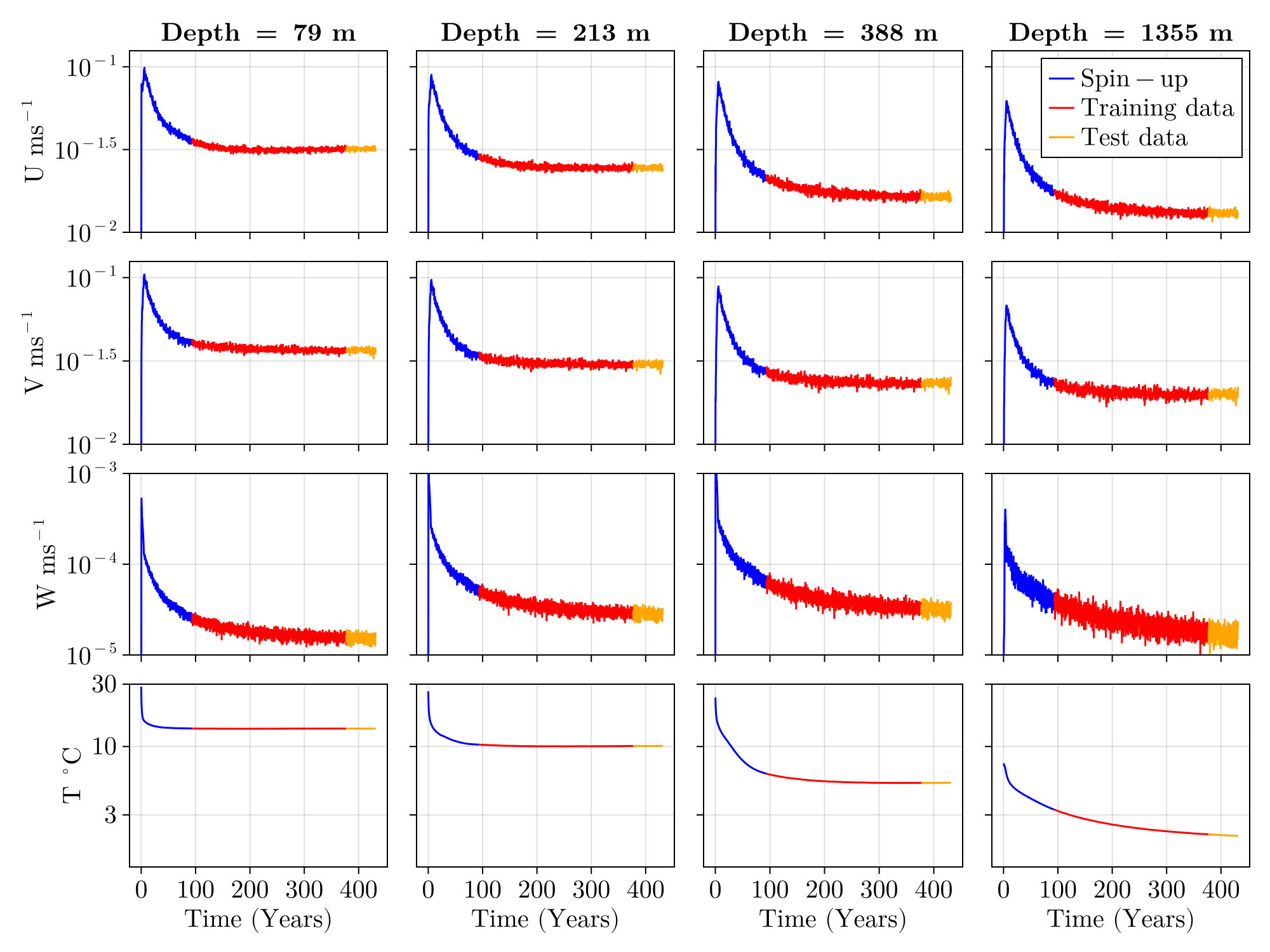}}
\caption{{Evolution of $L^1$ norms in time}. We show the average value of the absolute value of various fields (see Equation \ref{l1_error}) as a function of time in years on a logarithmic scale. The surface fields are in quasi-statistical equilibrium, whereas the fields near the bottom of the domain are still converging, especially for the temperature field. The different colors represent the discarded spin-up data (blue), training data (red), and test data (orange) for training a score-based diffusion model. }
\label{fig:non_equ}
\end{center}
\vskip -0.2in
\end{figure*}

Our simulation is not in statistical equilibrium for all observables of the flow, as can be observed from Figure \ref{fig:non_equ}, which shows the $L^1$ norm (the average absolute value) of all variables at a few selected depths, e.g.
\begin{align}
\label{l1_error}
\| V \|_1 =  \frac{1}{60^{\circ}} \frac{1}{60^{\circ}}\int_{15^{\circ}}^{75^{\circ}}  \int_0^{60^{\circ}}  |V(\lambda, \varphi, z, t) | \text{ d}\lambda \text{d}\phi.
\end{align}
We use this measure as it is sensitive to localized fluctuations, such as the root-mean-squared value. Including metric terms does not change the overall trend and is unnecessary for the present case, but we will include them later when calculating physically meaningful quantities rather than statistical measures of non-stationarity. The initial spinup period (blue) is discarded. Thus, all the data in the work uses the time periods associated with the middle portion (red) and the end portion (orange) of the simulation. Closer to the surface (first column), it appears that a quasi-statistical equilibrium has been reached, whereas at depth (last column), statistical equilibrium has not been reached, especially for temperature. This quasi-equilibrium nature of the ocean is a feature we wished to capture in our dataset when investigating the limits of statistical inference predicated on past data. We comment that the sea surface height's $L^1$ norm is also not in equilibrium. Stated differently, our ``data distribution" (red) will be different than our ``validation distribution" (orange) and will contribute as an extra source of error. \cblue{That being said, there are many statistics which can, for the most part, be regarded as being in statistical equilibrium such as the surface horizontal velocity components.}

\subsection{Probabilistic Models}
\label{s:prob_mod}

In the present work, we draw samples from subsurface fields, $U, V, W, T$, given the sea surface height $\eta$, where the underlying probability distribution comes from the data distribution generated by the time series of the primitive equations. Mathematically, this is represented as a conditional probability distribution, $\mathcal{P}$, 
\begin{align}
\label{basic_problem}
\text{conditional distribution} = \mathcal{P}(\text{subsurface fields} | \text{surface fields}).
\end{align}
The conditional probability distribution outputs four fields $U, V, W, T$ at every grid point in three-dimensional space, conditioned on the SSH $\eta$. Concretely, samples from this distribution produce $4 \times 256 \times 256 \times 15$ numbers conditioned on $256 \times 256$ numbers using the full resolution of the simulation. This high-dimensional conditional probability distribution is learned from the data set using deep-learning methods, see Section \ref{s:mcd}\ref{s:generative}.

If the conditional information is perfectly informative, there is a functional relationship, $\mathcal{F}$, between SSH and predicted fields and predicted fields. In this case, the conditional distribution collapses to a delta function, 
\begin{align}
\label{informative}
\mathcal{P}(\text{subsurface fields} | \text{surface fields}) = \delta(\text{subsurface fields} - \mathcal{F}[\cblue{\text{surface fields}}]).
\end{align}
If the provided information is ``not useful'' or maximally uninformative, then the conditional probability distribution, $\mathcal{P}$, should reduce down to the marginal distribution of the data, e.g., a random sample with statistics similar to those the generative model has been trained on independently of the surface information, 
\begin{align}
\label{useless}
\mathcal{P}(\text{subsurface fields} | \text{surface fields}) = \mathcal{P}_M(\text{subsurface fields}),
\end{align}
where $\mathcal{P}_M$ is the marginal distribution of the subsurface fields. Most inference tasks lie between these two extremes, Equations \ref{informative} and \ref{useless}. We comment that we use the word ``marginal'' as is commonly used in the statistics literature with regard to distributions. 

Traditional machine learning methods find a functional form $\mathcal{F}$ between input and outputs. This methodology is well-suited for problems where such a functional form exists. Otherwise, traditional methods that minimize a root-mean square (RMS) error to find a functional form can only infer the average value of the conditional probability in Equation \ref{basic_problem}, \cblue{that is 
\begin{align}
\label{trad_ML}
\mathcal{F}[\text{surface fields}] = \mathbb{E}_{\text{subsurface fields}}[\mathcal{P}(\text{subsurface fields} | \text{surface fields})],
\end{align}
much like a ``mean" statistic minimizes the RMS error of a random variable. (See Appendix A for a proof of this result.) If the conditional distribution is a delta-function, as in Equation \ref{informative}, little is lost by performing the average; however, if there is significant variance in the output field, then the best that a deterministic model can do is to average over such variance. At the extreme, the output of a deterministic model will produce the ensemble average of the previously seen data, e.g. the average value of the marginal distribution, Equation \ref{useless}. }
We surmise that this observation partially explains why fields drawn from a conditional distribution better represent the statistical properties of fluid flows compared with those learned via a deterministic functional relationship. Details will not be ``washed away" from averaging. 

\cblue{An important consequence of Equation \ref{trad_ML} is that we can compare generative models to traditional regression techniques indirectly by taking the ensemble average of the output given by the generative model. In this way we can see what details are missed by using non-generative approaches as well as provide a baseline for when we expect traditional approaches to outperform generative methods. This observation is independent of the neural network architecture and comes instead directly from the objective function that is being minimized by an ``RMS loss'' of a traditional method versus the ``denoising score-matching loss'' of a generative approach.  }

\subsection{Score-Based Diffusion Model}
\label{s:generative}
Our work leverages recent advancements in score-based generative modeling, particularly building upon foundational contributions, \citep{Song2019, Song2020Improved, Ho2020, useful_diffusion}. These methods allow for efficient sampling of arbitrary probability distributions, including conditional distributions.   

The key deep-learning methodologies to sample the conditional distribution include reverse diffusion and conditional denoising score-matching \citep{Anderson1982, Hyvarinen2005, Vincent2011, Batzolis2021}. Architecturally, we adopt the U-Net framework \citep{Ronneberger2015}, a well-established model utilized in generative modeling \citep{Deck2023, bischoff2024unpaired}.

The simulated SSH data is coarse-grained from its original resolution of $256 \times 256 \times 15$ to $128 \times 128 \times 15$ by averaging neighboring horizontal points to expedite the training of the generative model. This lower-resolution dataset loses few features from the original but dramatically expedites the training process since the entire dataset can be loaded into computer memory. We take monthly snapshots over the simulation period to ensure that our data is sufficiently decorrelated. 

A coarse-grained sea surface height is the conditional information provided to the model, which serves as the primary input for generating samples. We examine various SSH coarse-graining levels to determine the subsurface fields' predictability. Specifically, we average neighboring points by factors of two until only a single value for the SSH remains as input for the deep-learning model. That is to say, we progress from a 50 kilometer resolution to a 6,400 kilometer resolution in latitude. This procedure results in 8 candidate SSH ``inputs'', at any fixed moment in time due to the different resolutions. Each state, $U, V, W, T, \eta$, is separately normalized at each of the 15 levels of depth so that each field has the spatio-temporal mean removed and is divided by twice the spatio-temporal standard deviation. This rescaling keeps the fields roughly $\mathcal{O}(1)$ for training purposes. Thus, $8$ numbers are used separately to rescale $U, V, W, T$ at each of the 15 vertical level, and $2$ numbers are used to rescale $\eta$, leading to a total of $8 \times 15 + 2$ scaling factors. 

We train a single neural network that simultaneously outputs the entire 3D field for $U, V, W, T$  and takes in all coarse-grained inputs. We enable the latter step by randomly choosing among the 8 different coarse-grained SSH. We use a 2D UNet with 61 input channels, 60 output channels ($15 \times 4$ 2D output fields and $1$ 2D input field), and $\mathcal{O}(10^7)$ parameters. The resulting network has a $\mathcal{O}(10^2)$ megabytes memory footprint, which we compare to the $\mathcal{O}(10^2)$ gigabytes used to train the generative model. In this sense, we can think of the generative model as a statistical compression technique which, in the present case, reduces the storage by a factor of $10^3$. The Adam optimizer \citep{kingma2014adam} is employed for training the network. 


\section{Results}
\label{s:results}
Given the ocean simulation data (OcS) from Section \ref{s:mcd}\ref{s:ocean} and the neural network architecture from Section \ref{s:mcd}\ref{s:generative}, we perform statistical inference by drawing 100 sample fields using the reverse diffusion process, given SSH. We test the enural network by feeding SSH from the last 50 years of the double-gyre simulation which were not used during training. The predictions of the subsurface variables are tested for the various coarse-grained SSH input data.


The generative models draw many sample realizations from one SSH, yet we only have one ground-truth simulation to compare against for a given SSH. As such, we will look two ways of comparing the distributional output to the ``ground truth''. The first is to compare the ensemble mean of the generated samples to the ground truth since an ``ensemble mean'' is relatively easy to compute and serves as a good ``average" guess. This choice is most useful when the conditional information is sufficiently informative, as in the limit described by Equation \ref{informative}. The second is to simultaneously compare the ocean simulation output to all resulting ensemble members, which yields a probability distribution. We can then compare this probability distribution to one where we pick one of our generated samples (for example, the first) and compare it to all the others (for example, 2-100). We then check to see if the two probability distributions are consistent. This latter choice is a purely ``data-driven" way to determine how accurate the resulting predictions of the model are. A well-trained model (which requires sufficient data and an appropriate choice of architecture) makes meaningful estimates of the range of plausible fields for a given SSH, and thus the two resulting distributions should be similar. We will return to this point later.

\subsection{Baseline Performance of Generative Model}
\begin{figure*}[htbp]
\vskip 0.2in
\begin{center}
\centerline{\includegraphics[width=1.0\textwidth]{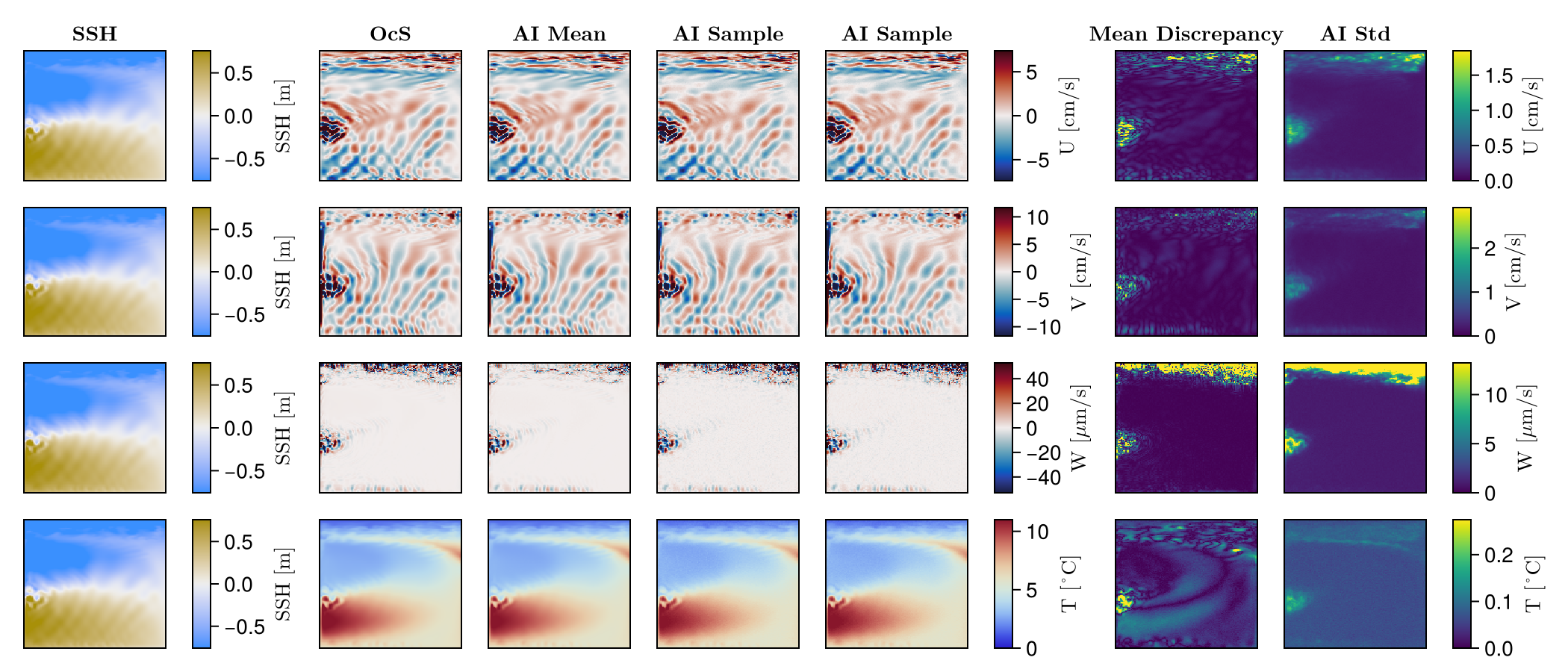}}
\caption{{Statistical inference of the ocean interior state from SSH.} Given the SSH input field (left column), we reconstruct the ocean state: the velocity component fields $U, V, W$, and temperature $T$ (second column). All fields are generated simultaneously. We show the mean over 100 samples predicted by the neural network (third column) and two representative samples (fourth and fifth column). The absolute difference between the AI mean and the OsC ground truth is displayed in the second-to-last column alongside the AI standard deviation in the last column.  }
\label{fig:predict_states}
\end{center}
\vskip -0.2in
\end{figure*}

We start by examining how well the SSH constrains the three components of velocity and temperature at a $400$ meter depth. We show the result of a trained neural network model for generating conditional samples in Figure \ref{fig:predict_states}. The sea surface height (first column) and corresponding simulated variables at a 400-meter depth (second column) are taken from 50 years in the future from the end of the training set. Each row corresponds to a different ocean state component, $(U, V, W, T)$,  as labeled. Using the generative model, we draw 100 samples from the conditional distribution with this sea surface height as the conditional information. We then take the ensemble average over these samples (third column). We show 2 of the 100 samples in the fourth and fifth columns to illustrate the ``generative'' nature of the prediction from the neural network. 

We see an overall agreement between the ground truth (second column) and the expected value of the model predictions (third column). This result means that a substantial ageostrophic component is also captured. The ensemble mean of the generated samples looks similar to the ground truth and the representative generated samples since there is little diversity in the generated samples. The conditional distribution is nearly a delta function, suggesting that the sea surface height is very informative of the ocean state and that a functional relationship can be found between the sea surface height and sub-surface currents. In other words, the present inference task is leaning towards Equation \ref{informative}. \cblue{Given that a traditional supervised learning task would produce the ensemble mean prediction of a generative model, low sample diversity would imply that a supervised learning task would perform well for the horizontal velocity components and the temperature field; however, the vertical velocity field has significant sample diversity (strong fluctuations in the northern region as can be seen by comparing the ensemble mean to the generated samples), implying that a traditional machine learning model would miss these fluctuations and ``average them out''. }

The sixth column shows the absolute value of the difference between the OcS and the AI mean, and the seventh column shows the ensemble standard deviation of the AI method. We represent predictive uncertainty by using the generated fields' standard deviation at each grid point.  The largest standard deviation is associated with the northern convecting region, especially for vertical velocity. We surmise that the sea surface height alone is insufficient to constrain the pointwise vertical velocity magnitude in these regions. \cblue{A traditional machine learning model would average out these fluctuations to produce the average vertical velocity field as given by the ``AI Mean'' column. We see the flexibility of the generative model in informing our ability to apply traditional machine learning models.} The second largest source of uncertainty is the eddying gyre region, which contributes to the uncertainty of the velocity components. The mean discrepancy column and the AI Std column have similar spatial structures, suggesting that the method can identify regions in which it is less capable of prediction.

\cblue{Although mean discrepancy and the standard deviations are similiar in structure and magnitude across the velocity fields, we see that the temperature field has a different spatial structure. Although the strongly eddying region and the northern convective remain the largest source of uncertainty there is also a uniform background uncertainty in the domain as well. This uniform uncertainty of the temperature field could perhaps be related to the non-stationarity of the domain average temperature field, which is a statistic that evolves on a much longer timescale than others in the OcS. Ultimately, the relative errors of the temperature field are smaller as compared to the velocity fields. The standard deviation of the generative model's temperature field overestimates this (small) uncertainty.}

\subsection{Ocean State Predictions and Uncertainties with Degraded Surface Height Input}

\begin{figure*}[htbp]
\vskip 0.2in
\begin{center}
\centerline{\includegraphics[width=\textwidth]{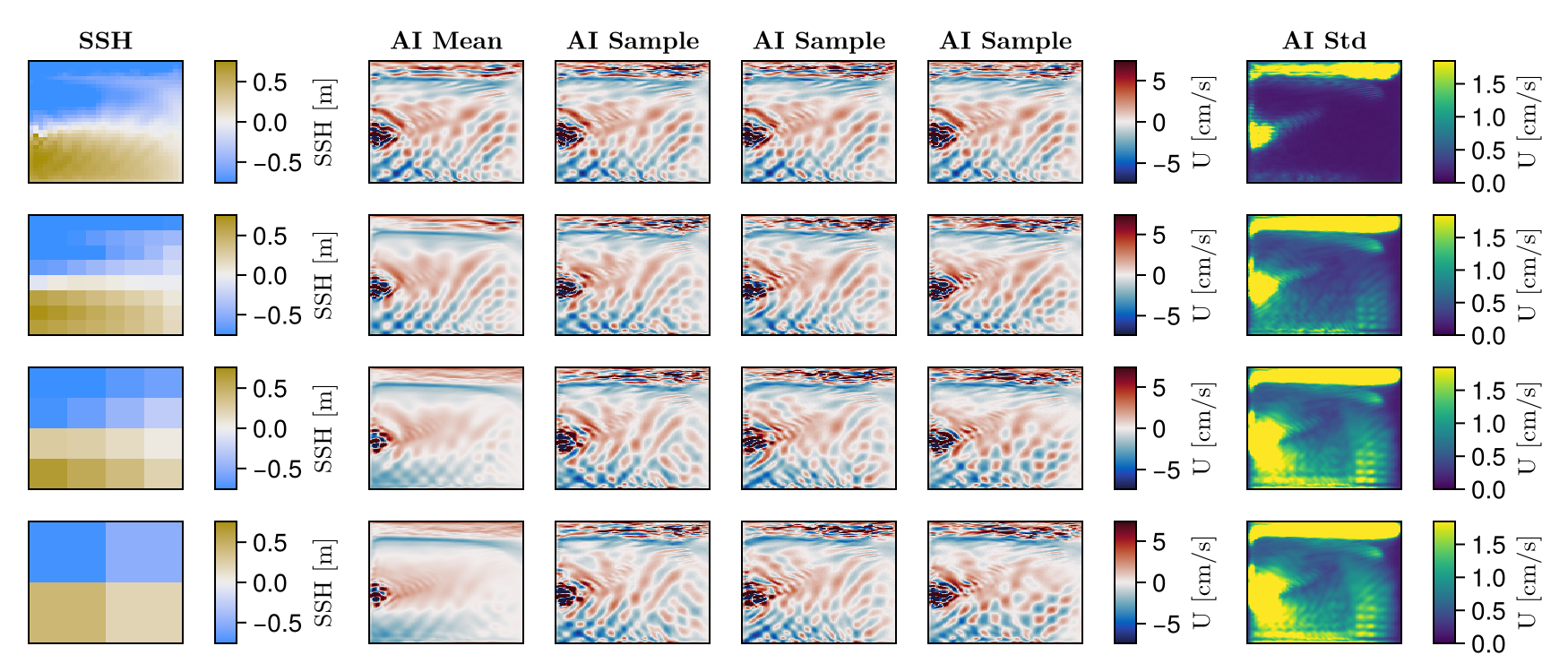}}
\caption{{Statistical inference of zonal velocity given the different resolutions of the free surface.} Given the sea surface height at several coarse-grained resolutions (first column), we show the output of the generative AI model ensemble in the subsequent columns. We show the average value over 100 samples predicted by the neural network (second column) and representative samples (third, fourth and fifth columns). The pointwise standard deviation of the generative AI ensemble is in the last column. The latitudinal resolutions of SSH are, from top to bottom, 200 kilometers, 800 kilometers, 1,600 kilometers, and 3,200 kilometers.   }
\label{fig:predict_coarse_grain_V}
\end{center}
\vskip -0.2in
\end{figure*}

\begin{figure*}[htbp]
\vskip 0.2in
\begin{center}
\centerline{\includegraphics[width=1.0\textwidth]{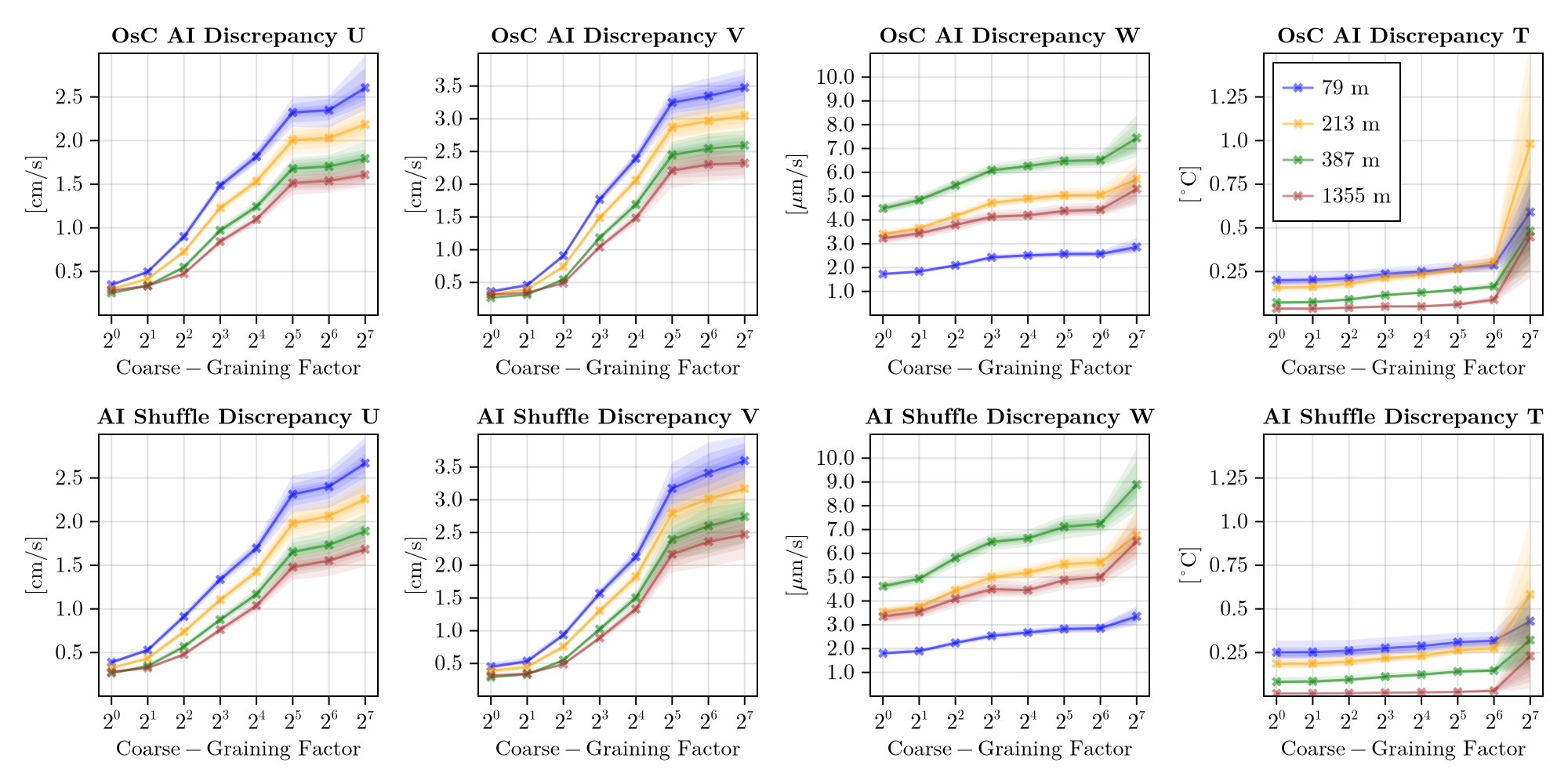}}
\caption{{Discrepancy as a function of depth and coarse-graining for a fixed SSH}. We show the distribution calculated by the difference in the neural network prediction 100 member ensemble and the ocean simulation with the $L^1$ norm in the first row. In the second row we show a purely AI calculation of error given by taking differences between ensemble members and using the $L^1$ norm. The columns represent the different fields, $U, V, W, T$. Within a panel the different colors represent different depths, and the horizontal axis corresponds to different coarse-graining factors of SSH with $2^0$ representing the full resolution (50 kilometers) and $2^7$ representing an average value of SSH (6,400 kilometer resolution).}
\label{fig:error}
\end{center}
\vskip -0.2in
\end{figure*}

Next, we check the ability to predict the ocean interior upon losing sea surface height information. We do this by applying a filter to the sea surface height where we average neighboring points and provide this filtered free surface as an input to the generative model. We use the same neural network model for this inference task as it was simultaneously trained on all coarse-grained fields. We show the resulting statistical inferences for the zonal velocity in Figure \ref{fig:predict_coarse_grain_V} and quantitative errors as a function of depth and coarse-graining in Figure \ref{fig:error}.

We first focus on Figure \ref{fig:predict_coarse_grain_V}.   In the first column, we show the sea surface height at a 200 kilometer resolution in latitude ($32 \times 32$ number of points) in the top, with a 800 kilometer resolution ($8 \times 8$ points) in the second row and coarse grain by a factor of two in each subsequent row to 1,600 kilometers and 3,200 kilometer resolutions in latitude, respectively. Although the network predicts all the velocity components and buoyancy at each vertical level, we restrict the visualization to the zonal velocity at a $400$ meter depth. The outputs can be directly compared to the first row of Figure \ref{fig:predict_states}.  The generative model produces 100 samples for each coarse-graining level. The second column shows the ensemble average zonal velocity, and columns three through five show randomly selected samples. We see by inspection that the ensemble average departs progressively the samples as we proceed downwards through the rows and that the samples become increasingly diverse. In addition, we show the point-wise standard deviation of the 100-member ensembles in the last column to quantify the diversity of the generated samples.

\begin{figure*}[htbp]
\vskip 0.2in
\begin{center}
\centerline{\includegraphics[width=\textwidth]{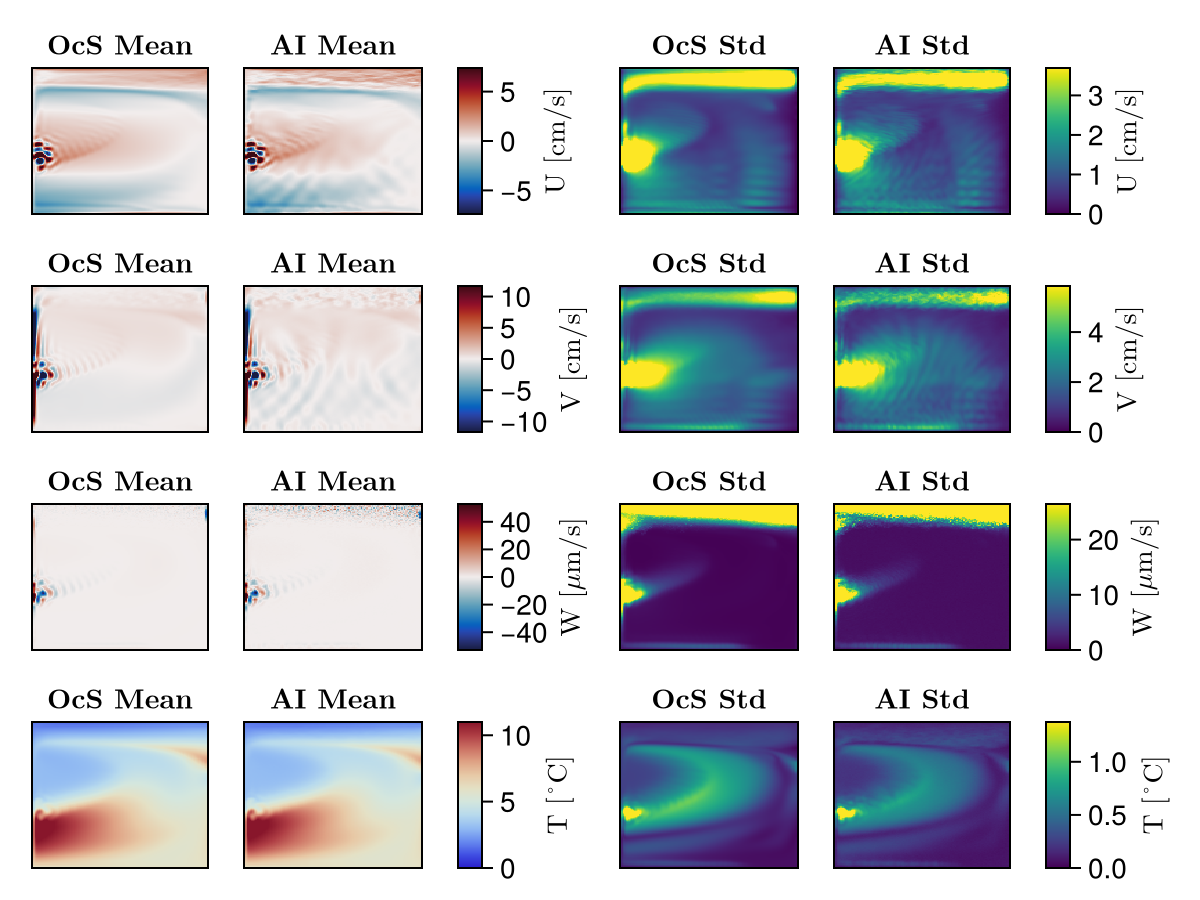}}
\caption{{Comparison of the AI ensemble mean to the OsC temporal mean.} Using fully coarse-grained SSH as inpute to the network, we generate 100 samples and calculate point-wise statistics. We compare these ensemble statistics to the temporal analog of the ocean simulation over the training period.}
\label{fig:marginal}
\end{center}
\vskip -0.2in
\end{figure*}

These observations suggest that an SSH coarse-grained beyond $\mathcal{O}(1,000)$ kilometers is no longer informative since the generated samples become so diverse as to reproduce statistics of the data distribution, a ``worst case'' scenario for the generative method.  As we coarse-grain the free surface (lose information), the inference task moves from something informative (similar to Equation \ref{informative}) towards the uninformative (similar to Equation \ref{useless}) marginal distribution of the training data. \cgreen{Taken together we see that the generative model can quantify the ``non-existance'' of a unique purely deterministic method for predicting subsurface states; the problem becomes inherently ill-posed from a deterministic perspective (i.e. trying to predict a single unique subsurface state is not possible, because the given information is insufficient to constrain potential subsurface states). For a deterministic method to succeed it must be supplemented with either additional information, some regularization or a criteria to select from the multitude of possible subsurface configurations, (e.g. choosing the most probable subsurface state, an ensemble mean subsurface state, or some other selection criteria).}

We compare the sampled distribution using the most uninformative (i.e. fully coarse-grained over the entire domain) SSH as the conditional information to the marginal distribution of the data over the training period in Figure \ref{fig:marginal}. This corresponds to the maximally uninformative case and the goal is to show that uninformative data reduces to the drawing samples similar to the training data. To do so we compare the ensemble statistics of the generative model to the temporal statistics of the training data. For brevity we use a fixed depth. The rows correspond to the different fields $U, V, W, T$ at a 400 meter depth. The OcS mean/std is a temporal mean/std over the training data set and the AI mean/std is generated from a 100 member ensemble using only the domain average free-surface height as input, i.e. we remove spatial structure from the input field. We see that there are still wave-like oscillations present in the AI predictions which we attribute to insufficient averaging from the limited generated samples since using less ensemble members exacerbated the oscillations, but overall there seems to be good agreement between the two calculations. 

In Figure \ref{fig:error}, we quantify the discrepancy between the ensemble average of the AI prediction and the ocean simulation in predicting the various fields as a function of resolution and depth. We take the difference between the OcS simulation and each ensemble member individually. We use the $L^1$ norm (average absolute error) to illustrate the discrepancy and the resulting distribution with different bands whose contours represent the $0.6, 0.7, 0.8$, and $0.9$ quantiles, from darker to lighter. The points are the ensemble average of the resulting distribution. We denote this as the ``OsC AI Discrepancy''. We also perform an analogous computation purely from the AI ensemble, which we call the ``AI Shuffle Discrepancy'' in the second row. Here, we plot the distribution given by the difference between different ensemble members. Specifically, we take the $L^1$ norm between ensemble members 1 and 2, 2 and 3, 3 and 4, and so forth to ensemble members 100 and 1, yielding $100$ samples of an $L^1$ norm, which we then take to be a distribution to compare against the distribution given by the ``OsC AI Discrepancy''. A similar calculation  is to take ensemble member $1$ as a ``fake'' ground truth, and take the difference between the other ensemble members $2$ through $100$. We choose the ``shuffled'' version to reduce variation due to a ``unlucky'' choice of ground truth ensemble member. 

We see that (for all fields), as we decrease the resolution of the sea surface height, the discrepancy between generated fields and the OsC increases. The average discrepancy of the $U$ and $V$ velocity field components decreases as a function of depth since their magnitudes decrease with depth. The temperature field $T$ has the opposite trend, with the largest average discrepancies coming from the surface, perhaps due to the larger overall temperature values near the ocean's surface. The vertical velocity exhibits a less clear trend, with the largest average discrepancy coming from a 400-meter depth and the lowest coming from the near-surface value. A potential explanation lies in the fact that $w = 0$ at the surface and the seafloor, so the largest value should come from somewhere at mid-depth. 

The trends of the first row are quantitatively reflected in the second row. The ``Shuffle Discrepancy'' is similar to the difference between the simulation output and the AI ensemble. This observation suggests that the AI method can provide meaningful uncertainty estimates in so far as the data it was trained on is similar in distribution to what is seen in the future. For example, we see a difference between the 1355-meter depth ``OsC AI'' temperature discrepancy and the corresponding ``AI Shuffle Discrepancy'' where the AI Shuffle Discrepancy underpredicts the associated error. We attribute this to the lack of equilibrium of the ocean simulation depths, as reflected in Figure \ref{fig:non_equ}. 

\begin{figure*}[htbp]
\vskip 0.2in
\begin{center}
\centerline{\includegraphics[width=\textwidth]{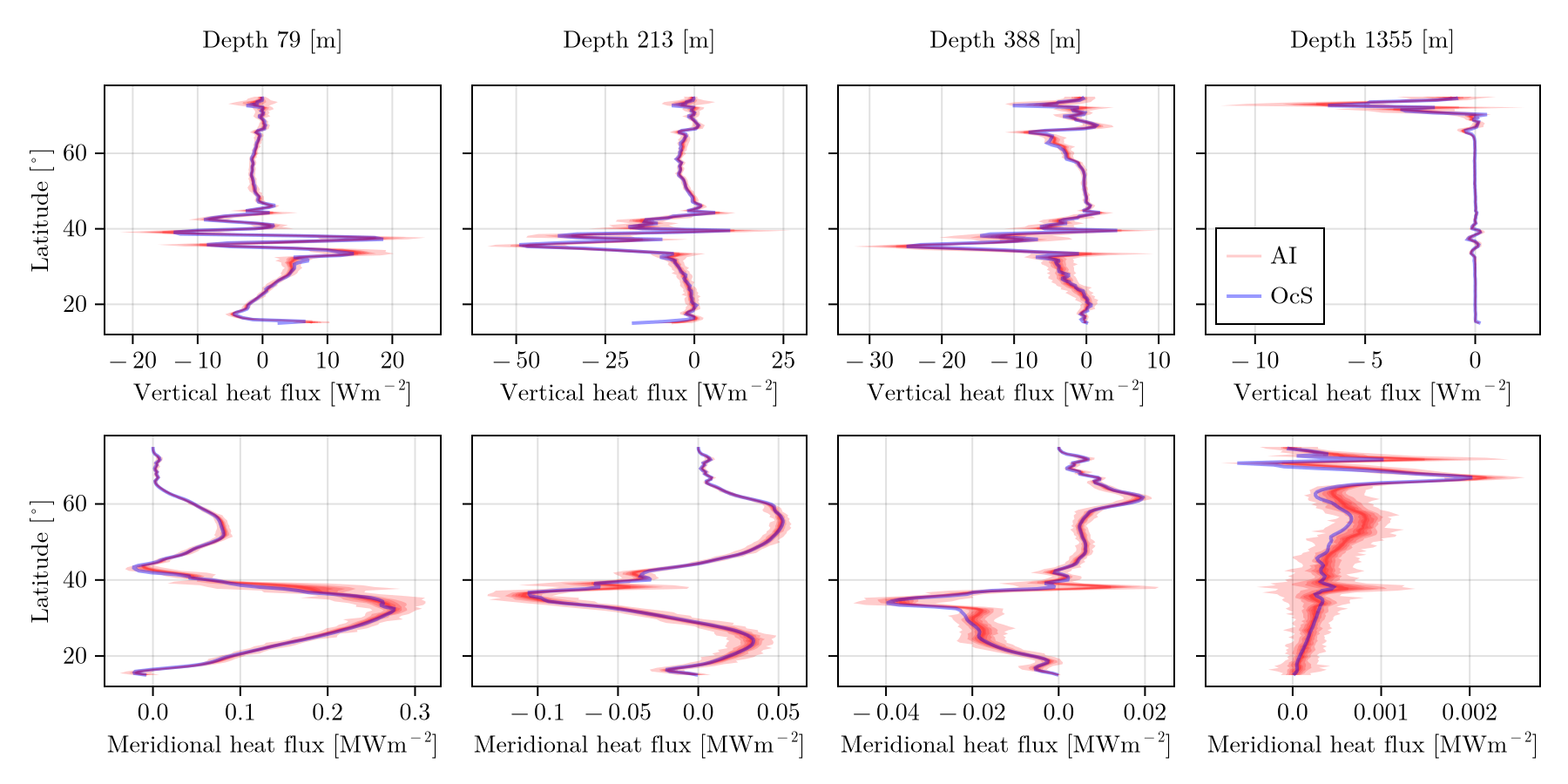}} 
\caption{{Zonal average of vertical and horizontal eddy heat flux for a fixed SSH at various depths.}  We show the zonal average of the vertical (top row) and meridional eddy heat flux (bottom row) over the domain for each latitude at a fixed moment time and various depths in blue. The neural network results using the full-resolution sea surface height for input are in red, with the solid red line representing the ensemble average heat flux. The ribbons represent the $0.6, 0.7, 0.8$, and $0.9$ quantiles, from darker to lighter, as computed with the 100-member ensemble. Each column shows a different depth, and the relative uncertainty (red ribbon region) increases as we go deeper into the domain.  }
\label{fig:avg_hf}
\end{center}
\vskip -0.2in
\end{figure*}

\subsection{Heat Flux Predictions and Uncertainties with Degraded Surface Height Input}
Thus far, we have only examined statistics within a given field but have not looked at correlations across fields. We address this by calculating the vertical and meridional eddy heat flux. We do not show the full heat flux because its value depends on arbitrary temperature units. Also, we use the zonal averages of the fluxes for illustrative purposes. Mathematically, we use 
\begin{align}
    \overline{\cdot} \equiv \frac{1}{60^{\circ}}  \int_0^{60^{\circ}} &\cdot d \lambda ,  \\
    \text{vertical eddy heat flux} &= \cos(\varphi) \left( \overline{wT} - \overline{w}\overline{T} \right) , \\
    \text{meridional eddy heat flux} &= \cos(\varphi) \left( \overline{vT} - \overline{v}\overline{T} \right).
\end{align}

\begin{figure*}[htbp]
\vskip 0.2in
\begin{center}
\centerline{\includegraphics[width=\textwidth]{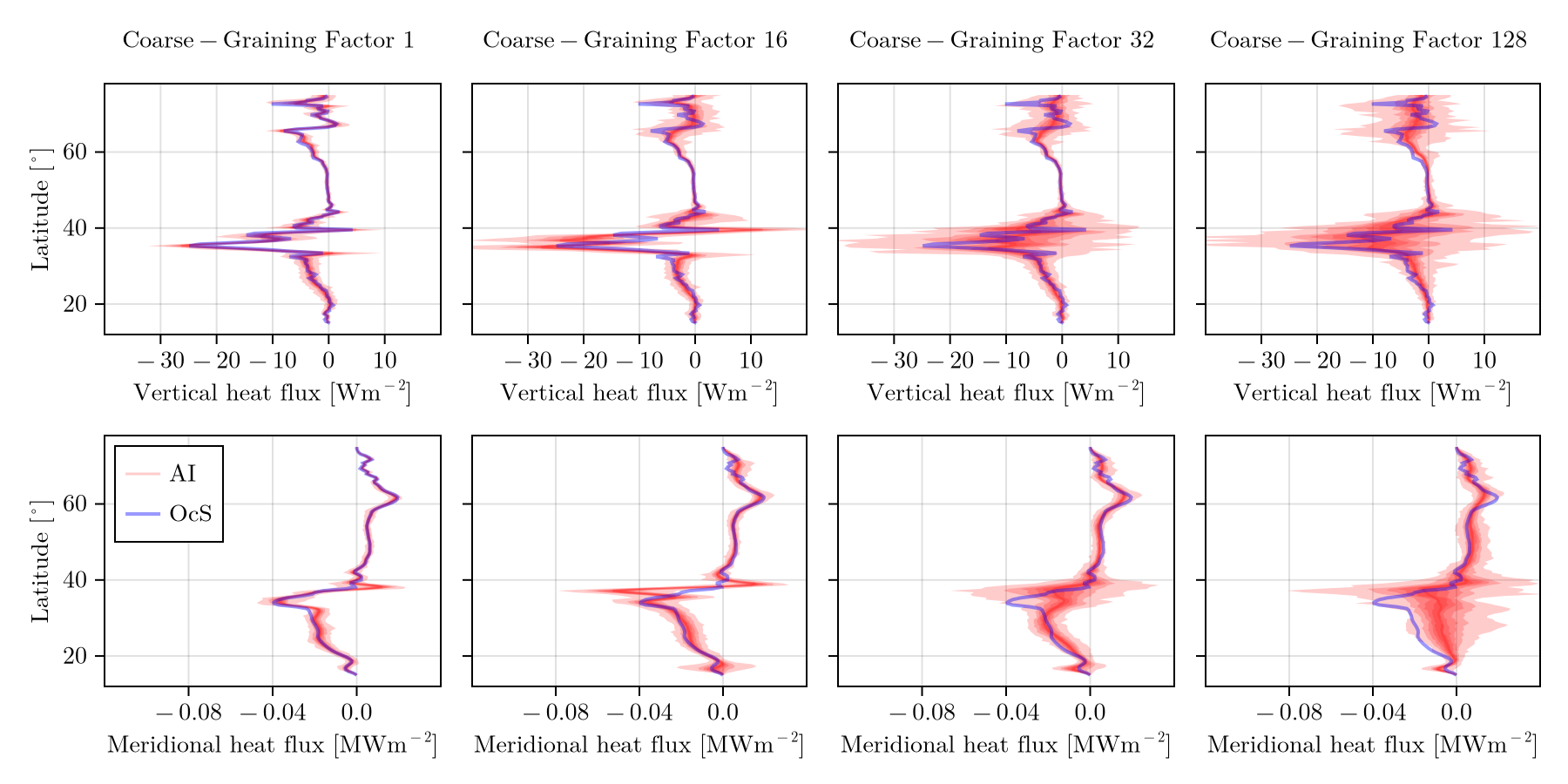}}
\caption{{Zonal average of vertical and horizontal eddy heat flux for a fixed SSH at a 400 meter depth.} We show the zonal average of the vertical (top row) and meridional heat flux (bottom row) over the first fifteen degrees of the domain for each latitude at a fixed moment time and a depth of 400 meters in blue. The neural network results are in red, with the solid red line representing the ensemble average heat flux. The ribbons represent the $0.6, 0.7, 0.8$, and $0.9$ quantiles, from darker to lighter. Each column uses different levels of coarse-grained sea surface height. As the amount of coarse-graining increases, the confidence bounds (as represented by the red region) increase and reduce to the marginal distribution of the data distribution). Information about the location of the meander is lost when going from a factor of 8 reduction to a factor of 16 in reduction of resolution. See the first column and third row of Figure \ref{fig:predict_coarse_grain_V} for the corresponding SSH. }
\label{fig:avg_hf_degrade}
\end{center}
\vskip -0.2in
\end{figure*}

We show the heat fluxes as a function of latitude at various depths (in each column) in \cpurple{Figure \ref{fig:avg_hf}} for the highest resolution SSH. The ``ground truth`` results are the blue lines, and the generative AI results with 100 samples are shown as ribbon plots that capture the $0.6, 0.7, 0.8, 0.9$ quantiles from darker to lighter shadings of red. The top row is the vertical heat flux, and the bottom row is the meridional heat flux, both as a function of latitude. Generally, we see that the generative model predicts the heat flux well for all cases, with the ribbons increasing in relative size with depth: the meridional heat flux at the lowest shown depth is not (relatively) as well constrained.

We also calculate the eddy heat fluxes as a function of SSH resolution at a $400$ meter depth in Figure \ref{fig:avg_hf_degrade}. Similar to the previous plot, the top row is the vertical heat flux, the bottom row is the meridional heat flux, the ``ground truth'' results are the blue lines, and the generative AI results with 100 samples are shown as a ribbon with $0.6, 0.7, 0.8, 0.9$ quantiles from darker to lighter shadings of red. The heat flux prediction grows in uncertainty as we coarse-grain the sea surface height. The loss of information here is understood upon examining the first column of Figure \ref{fig:predict_coarse_grain_V}. As the sea surface height (SSH) field is increasingly coarse-grained, we lose precise information about the location of the jet separation point — the region where the western boundary current detaches from the wall and begins to meander eastward. This separation point is critical in organizing mesoscale eddy activity and lateral heat transport. When spatial detail in the SSH is removed, the model can no longer resolve where the jet begins to meander, leading to increased uncertainty in the predicted eddy heat fluxes, particularly in the downstream region where the current interacts with ambient stratification. \cblue{The ability of the generative model to produce plausible eddy fluxes in the ill-posed regime is another feature of generative models: a deterministic model would be relegated to making an approximation of the form $\mathbb{E}[v T] \approx \mathbb{E}[v] \mathbb{E}[T]$ where the expected value on the right hand-side is the output of the deterministic model's prediction for $v$ and $T$. This occurs for the same reason mentioned prior that a traditional machine learning model can (at best) produce the ensemble mean of what a generative model can produce. }

\begin{figure*}[htbp]
\vskip 0.2in
\begin{center}
\centerline{\includegraphics[width=1.0\textwidth]{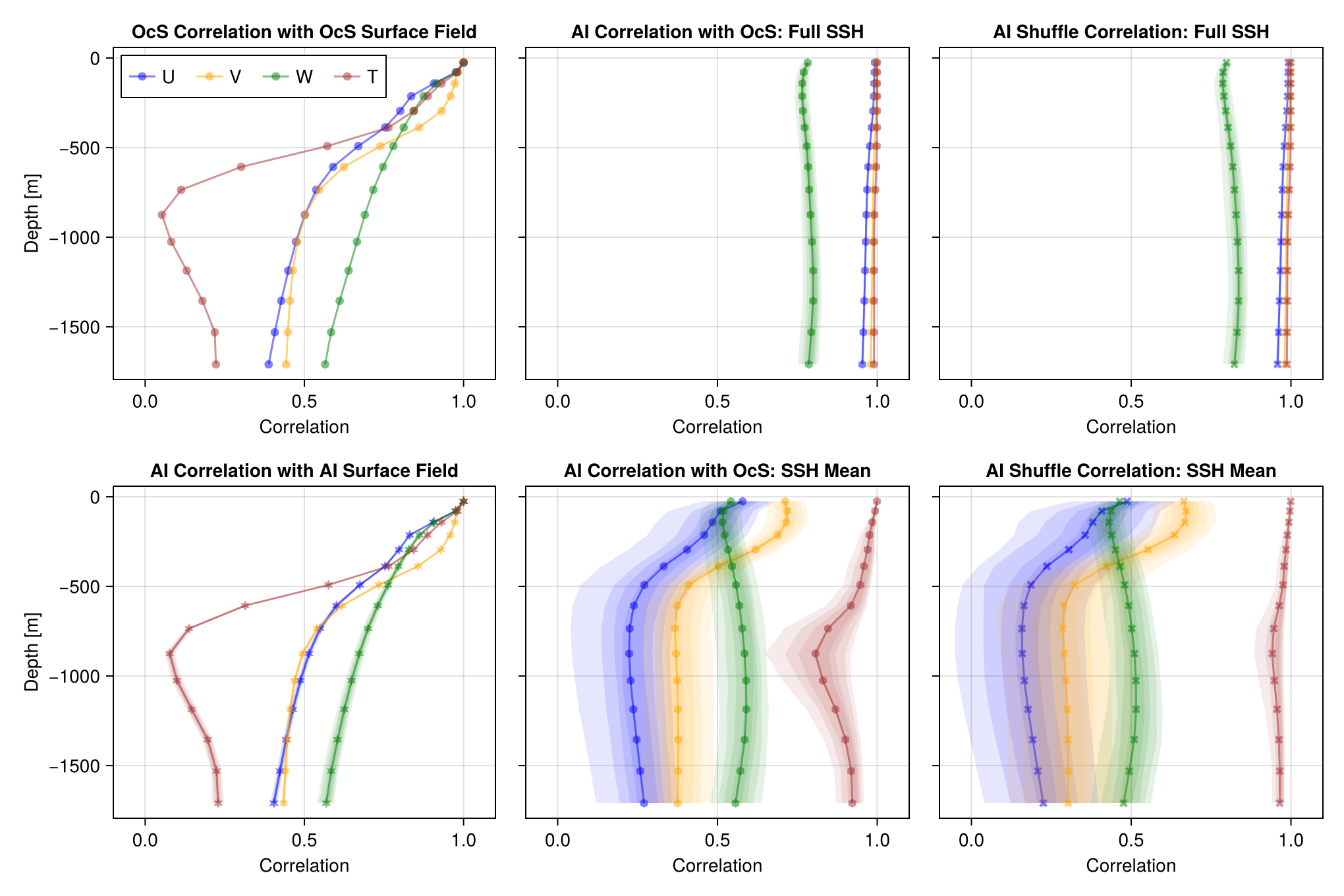}}
\caption{Correlation coefficients. Left column: correlations with the surface field in OcS (top) and the generative AI outputs (bottom). Middle column: cross-correlations between AI-predicted fields and OcS profiles, using full-resolution SSH (top) and domain-averaged SSH (bottom) as conditioning inputs. Right column: ensemble-based correlations computed purely from the AI model for full-resolution SSH (top) and domain-averaged SSH (bottom).}
\label{fig:depth_correlations}
\end{center}
\vskip -0.2in
\end{figure*}

\begin{figure*}[htbp]
\vskip 0.2in
\begin{center}
\centerline{\includegraphics[width=1.0\textwidth]{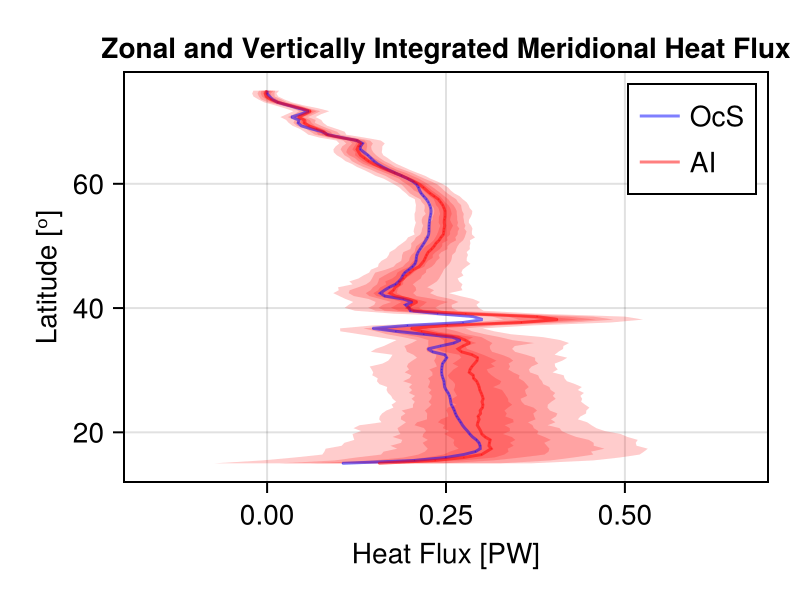}}
\caption{{Zonal and vertically averaged heat transport.}  The instantaneous transport as a function of latitude is estimated from the model simulation (blue line) and with the AI model using only sea surface height information (red line). The pink shading represents the $0.6, 0.7, 0.8$, and $0.9$ quantiles, from darker to lighter.}
\label{fig:int_x_z_hf}
\end{center}
\vskip -0.2in
\end{figure*}

\subsection{Statistics Combining Various Depths}
Understanding how well generative models predict subsurface ocean dynamics at different depths is crucial for assessing their viability in state estimation and climate modeling. Next we assess the generative model's ability to aggregate statistics across different vertical levels. Specifically, to evaluate the model's performance, we compare the predictions of the generative model against the ocean simulation for correlations across depths, the zonally and vertically integrated meridional heat flux, and the meridional overturning stream function in depth coordinates. We do this for the same 50-year future SSH used in previous sections.

Figure \ref{fig:depth_correlations} quantifies the vertical correlation of the subsurface fields under various inference settings. Here we use the correlation coefficient $\mathcal{C}$. This quantity for the meridional velocities at two different depths $V_1$ and $V_2$ is calculated as 
\begin{align}
\mathcal{C}(V_1, V_2) = 
\frac{\mathbb{E}[(V_1 - \mathbb{E}V_1)(V_2 - \mathbb{E}V_2)]}{\sqrt{\mathbb{E}[(V_1 - \mathbb{E}V_1)^2]} \, \sqrt{\mathbb{E}[(V_2 -\mathbb{E}V_2)^2}}
\end{align}
where the expected value is over the latitude-longitude points (i.e., a horizontal average without the metric terms), at a fixed depth, and for a fixed ensemble member (when appropriate). A similar calculation is performed for other quantities. For ensemble-based comparisons, results are shown as ribbon plots, with shading corresponding to the 0.6, 0.7, 0.8, and 0.9 quantiles (from dark to light).

The left column of Figure~\ref{fig:depth_correlations} shows the correlation between each field and its surface value, computed from the OcS simulation (top) and the generative AI model (bottom). These panels confirm that the generative model reproduces the vertical coherence present in the simulation. Furthermore, we note that the simulation deviates from equivalent barotropic behavior due to the changing correlation coefficient. The middle column compares AI-generated fields to the OcS profiles across depth, conditioned on either full spatial-resolution SSH (top) or mean (i.e. no spatial structure) SSH (bottom). The stronger correlations in the top row indicate that the fine-scale structure in SSH provides a meaningful constraint on subsurface variability. The existence of \textit{some} level of correlation in the lower row reflects the similarity between past ocean states and future ocean states as given by the simulation. After all the fields are not ``arbitrary'' but rather related to the past flow states of the model, e.g. regions in the past with no flow will remain with no flow. The right column presents ensemble-based correlations among AI samples conditioned on identical SSH inputs (top: full-resolution; bottom: no-resolution), serving as a ``shuffle test'' analogous to the discrepancy analysis in Figure~\ref{fig:error}. These correlations largely mirror the patterns seen in the middle column, indicating internal consistency in the generative model's outputs. A notable exception is seen in the temperature field at depth, where correlations drop significantly—a result we attribute to the lack of statistical equilibrium in the abyssal ocean.

We show in Figure \ref{fig:int_x_z_hf} the generative AI's ability to reproduce the zonally and vertically integrated meridional heat flux, 
\begin{align}
    \text{Meridional Heat Flux} = \rho c_p  R \cos(\varphi) \int_{-1800~m}^{0~m}\int_{0^{\circ}}^{60^{\circ} } V(\lambda, \varphi, z, t) \, T(\lambda, \varphi, z, t) \,\,\, {\rm d} \lambda {\rm d}z.
\end{align}
where $R$ is the radius of Earth, $c_p$ is the heat capacity of water, and $\rho_0$ is the reference density. The red line (ensemble mean of the generative AI prediction) and the resulting (from dark to light) ribbons represent the $0.6, 0.7, 0.8, $ and $0.9$ quantiles ribbons correctly encapsulate the ``ground truth" ocean simulation prediction. This quantity is evaluated 50 years in the future, similar to the previous sections. The uncertainty is quite large, yet the ensemble mean prediction is similar to ground truth. The uncertainty can likely be reduced by improving the model by better accounting for the vertical structure. Furthermore, the generative method need not be non-divergent, which may affect the resulting statistics. 

\begin{figure*}[htbp]
\vskip 0.2in
\begin{center}
\centerline{\includegraphics[width=1.0\textwidth]{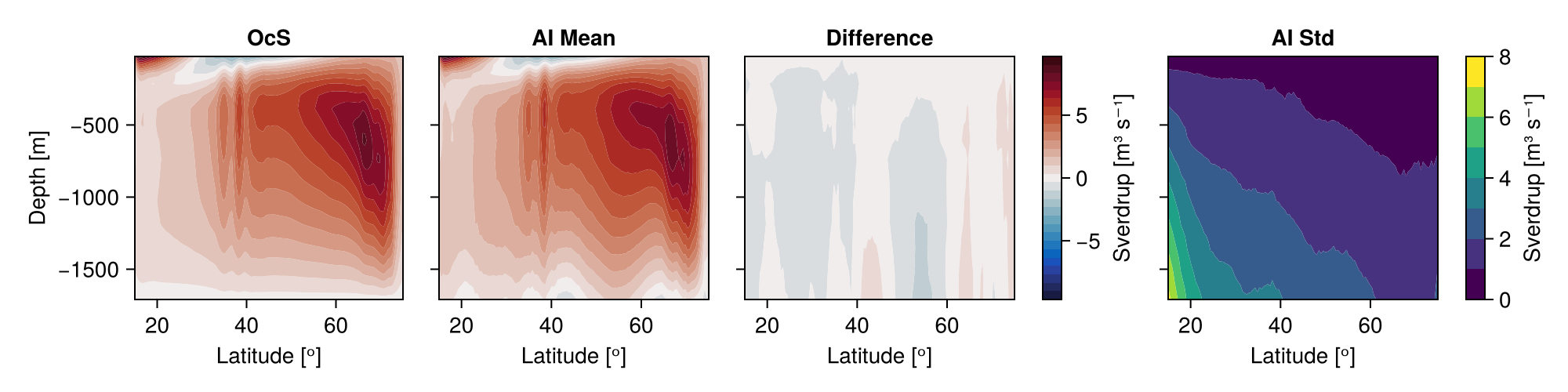}}
\caption{{The instantaneous overturning stream function in depth coordinates}. Here, we show the overturning stream function at each vertical level z and latitude for a fixed moment in time. The ocean simulation is compared to the AI mean. The difference between the ocean simulation and the AI mean is shown in the third panel and the AI standard deviation in the last panel.}
\label{fig:moc_z}
\end{center}
\vskip -0.2in
\end{figure*}

Lastly, we calculate the meridional overturning stream function, 
\begin{align}
\psi(\varphi, z, t) \equiv R \cos(\varphi) \int_{z}^0 \int_{0^{\circ}}^{60^{\circ}} V(\lambda, \varphi, z', t) d\lambda dz'
\end{align}
Figure \ref{fig:moc_z} shows the resulting overturning stream function calculation. The top left figure is the ocean simulation output, the middle top figure is the mean of the 100 generated samples, and the top right is the difference. The bottom left and bottom middle panels are arbitrarily chosen samples, and the AI uncertainty (bottom right) is the standard deviation of the 100 generated samples. 

We see that the overall structure in the upper 400 meters of the domain is well captured, with errors increasing with depth. Furthermore, we see a larger discrepancy at lower latitudes and mid-depth. The largest standard deviation of the AI ensemble is at low latitudes and at depth. The lack of statistical equilibrium contributes to the error here. The largest error comes from the pool of cold water slowly finding its way along the seafloor, from northern latitudes towards the lower latitudes. This process takes time, and was not shown to the neural network during training. Regardless, the resulting variability is overestimated.

\section{Conclusions}
\label{s:end}
This study has demonstrated the potential of generative modeling as a statistical framework for inferring the interior state of the ocean from surface observations. We explored the ability of diffusion-based generative models to reconstruct subsurface velocity and buoyancy field statistics using only information about the free surface in a statistically non-equilibrated system, offering a probabilistic approach to state estimation. Our key findings can be summarized as follows:

\begin{enumerate} 
\item  We have demonstrated, in an idealized setting, that ocean state estimation can be formulated as a conditional generative modeling problem, where score-based diffusion models sample from the distribution of subsurface velocity and buoyancy fields given surface data. This approach captures inherent uncertainties in subsurface reconstructions, providing plausible subsurface fields consistent with the given information.  
\item The generative AI method accurately reconstructs key ocean circulation features—including mesoscale eddies, zonal jets, eddy-heat flux, and overturning circulations—across different depths, dynamical regimes, and surface data resolutions. Notably, the generative approach provided meaningful uncertainty estimates, particularly in eddying and convective regions, and correctly estimated its own predictive inaccuracy. 
\item While the model performs well in regions where surface observations strongly constrain subsurface dynamics, its predictive skill diminishes at greater depths and in areas where the ocean state is out-of-sample from ocean states similar to what has been seen previously. 
\end{enumerate}

Our results demonstrate that generative models might capture the ocean interior's statistical variability, even when conditioned on coarse-resolution surface data. However, decreasing surface observations' spatial resolution or increasing inference depth diminishes predictive skill. In the limiting case, when the surface signal is insufficient to constrain the interior, the generative model reverts to sampling from the marginal distribution of the training data. As shown in Figure~\ref{fig:error}, the most pronounced deterioration in performance occurs when the surface resolution is degraded from 100km (coarsening factor $2^2$) to 200km (
$2^3$), particularly for the horizontal velocity components where the eddy length scale is on the order of 100km. For temperature, predictive accuracy declines when large-scale gradient information is lost. These findings suggest that improvements in satellite resolution—from current AVISO-class missions (200km) to SWOT (50km)—could significantly enhance our ability to infer subsurface ocean dynamics \citep{aviso, swot}. Incorporating additional dynamical priors, such as quasi-geostrophic constraints or hybrid physics-informed machine learning architectures, may further improve the fidelity and robustness of generative reconstructions.

Future work will extend this approach to more complex ocean simulations that include seasonal cycles, bathymetry, and finer-scale turbulence, providing additional constraints on the generative model. Furthermore, integrating observational datasets with multiple modalities such as sea surface salinity, Argo float profiles, sea surface temperature, and velocity measurements from moorings will enable direct evaluation of the model’s performance in real-world ocean conditions, \citep{martin2025genda}. \cblue{Sparse measurements can also be incorporated as a part of the generative framework \cite{babu2025guidedunconditionalconditionalgenerative}. To properly incorporate satellite altimetry would be a space-time version of the ``in-painting'' exercise of that work, leading to a 4D probabilistic reconstruction.}

Measurement uncertainty must also be accounted for. One avenue for doing so would be to draw from a distribution of consistent free-surface heights, effectively sampling from $\mathcal{P}(\text{subsurface} | \text{surface}) \mathcal{P}_M(\text{surface})$ where $\mathcal{P}_M(\text{surface})$ incorporates uncertainty in the surface field.  Incorporating theoretical predictions as additional ``input'' for the generative model will also be explored. This way, we can use a theory as a ``physics-informed'' guess, further refined through the generative method and a numerical ocean simulation with higher fidelity physics. \cblue{For example, one could use theory to provide further conditional information to augment the prediction of the generative method. Symbolically we would be drawing from the distribution $\mathcal{P}(\text{subsurface fields} | \text{surface fields}, \text{theory}),$  
which could possibly further reduce ensemble spread in samples. }

The methods developed in this study contribute to the broader effort of integrating machine learning with physical modeling in climate science. By leveraging advances in generative AI and ocean state estimation, this work provides a foundation for a new approach to data assimilation, parameterization, and statistical inference in ocean modeling. Such methodologies can also be applied to atmospheric geophysical models as well as Earth system models to provide a probabilistic view of multi-scale dynamics. 

\clearpage
\acknowledgments
This work acknowledges support by Schmidt Sciences, LLC, through the Bringing Computation to the Climate Challenge, an MIT Climate Grand Challenge Project. We would also like to acknowledge various discussions with Milan Kl\"ower, Ludovico Giorgini, Mengze Wang, Keaton Burns, and Themis Sapsis on generative modeling and discussions with Carl Wunsch, Gregory Wagner, Abigail Bodner,  John Marshall, and Fabrizio Falasca on ocean state inference. We would also like to acknowledge the help of Mason Rogers. TB acknowledges the community at Livery Studio that provided a stimulating research environment.
%
%
\datastatement
Scripts for reproducing the figures and the neural network training are located at 
\url{https://github.com/sandreza/DoubleGyreInference}. The Oceananigans script for producing simulations is located at \url{https://github.com/CliMA/Oceananigans.jl/blob/as/z-star-coordinate/double_gyre_5.jl}. We developed a Jax code for implementing the deep learning method \url{https://github.com/sandreza/JaxDiffusion}.  

%






%



\appendix 

\appendix[A]
\label{s:mse_proof}
\appendixtitle{Proof of Conditional Expectation}

In the main text it was claimed that the minimizer of the mean-square-error (MSE) is the expectation of the a conditional distribution. Here we provide a standard proof of this claim for completeness. 

Minimizing the MSE of a dataset is equivalent to minimizing the cost functional $\mathcal{C}$,
\begin{align}
\mathcal{C}[\mathcal{F}] = \int \mathcal{P}_J(\text{subsurface fields}, \text{surface fields}) \| \mathcal{F}(\text{surface fields}) - \text{subsurface fields} \|^2
\end{align}
across all possible functions $\mathcal{F}$. Here $\mathcal{P}_J$ is the ``data-distribution'', i.e. the empirical distribution that consitutes the datapairs $(\text{surface fields}, \text{subsurface fields})$, as given by the numerical simulation. For brevity we will use ``$x = \text{surface fields}$'' and ``$y = \text{surface fields}$'' and $x^n$ to denote the surface field at timestep $t_n$ and $y^n$ to denote the subsurface field at timestep $t_n$. We denote the total number of timesteps in the dataset as $N$. With these abbreviations we have
\begin{align}
\mathcal{P}_J(x, y) = \frac{1}{N} \sum_{n=1}^N \delta(x - x^n) \delta(y - y^n)
\end{align}
as our explicit data-distribution, where $\delta$ is the Dirac-delta function; however, a well-trained neural networks will regularize the above data-distribution so that it may generalize to unseen data-pairs. Ideally, the regularized distribution should be related to the ``true distribution'' of the underlying numerical simulation. 

Our cost functional with the same notation is 
\begin{align}
\mathcal{C}[\mathcal{F}] = \int \mathcal{P}_J(x, y) \| \mathcal{F}(x) - y \|^2.
\end{align}
To minimize the above functional across all functions $\mathcal{F}$ we take the variational derivative of the above expression with a small function perturbation $\delta \mathscr{F}(x)$ and neglect higher order terms to yield
\begin{align}
\delta \mathcal{C} = \mathcal{C}[\mathcal{F} + \delta \mathscr{F}]- \mathcal{C}[\mathcal{F}] \approx 2 \int \mathcal{P}_J(x, y) \delta \mathscr{F}(x) \cdot \left(\mathcal{F}(x) - y \right) 
\end{align}
We now use the identity $\mathcal{P}_J(x, y) = \mathcal{P}_M(x) \mathcal{P}(y | x)$ to rewrite the above expression as 
\begin{align}
\int \mathcal{P}_J(x, y) \delta \mathscr{F}(x) \cdot \left(\mathcal{F}(x) - y \right) =\int dx   \mathcal{P}_M(x) \delta  \mathscr{F}(x) \cdot \int dy  \mathcal{P}(y | x)   \left(\mathcal{F}(x) - y \right). 
\end{align}
Carrying out the integral over $y$ we identify 
\begin{align}
\int dy \mathcal{P}(y | x)  \mathcal{F}(x) = \mathcal{F}(x) \text{ and } \int dy \mathcal{P}(y | x) y = \mathbb{E}_{y} \mathcal{P}(y| x),
\end{align}
which reduces our functional derivative to 
\begin{align}
\delta \mathcal{C}= 2 \int dx   \mathcal{P}_M(x) \delta  \mathscr{F}(x) \cdot \left(\mathcal{F}(x) - \mathbb{E}_{y} \mathcal{P}(y| x) \right).
\end{align}
Setting $\delta \mathcal{C} = 0$ and noting that this must hold across all variations $\delta \mathscr{F}$ yields the expression
\begin{align}
    \mathcal{F}(x) = \mathbb{E}_{y} \mathcal{P}(y| x),
\end{align}
i.e., the minimizer is the conditional expectation. 

At this point it is worth contrasting the above expression with what is obtained by a generative method. The generative model directly tries to capture (and sample from) the conditional distribution $\mathcal{P}(y|x) $ whereas the deterministic method calculates directly the expected value $\mathcal{F}(x) = \mathbb{E}_{y} \mathcal{P}(y| x)$. Thus to compare generative models to traditional models (minimizing with respect to MSE) we generate samples from the distribution $\mathcal{P}(y| x) $ and take an empirical average afterwards. 

There are, however, practical issues that may prevent such an identification of traditional machine learning methods with generative methods. One issue comes from different architectures or forms of implicit regularization that are used when training neural networks. Another issue is that the one may require a large number of empirical samples to generate a particular statistic. If either model overfits on the training dataset, then these results do not apply since we the connection between the two only happens for the ``global" minimizer of the cost functional for the underlying data-distribution (as opposed to the empirical data-distribution).

\bibliographystyle{ametsocV6}
\bibliography{references}

\end{document}